\documentclass[sigconf,screen,nonacm]{acmart}

\AtBeginDocument{%
  }

\usepackage{booktabs}
\usepackage{microtype}
\usepackage[utf8]{inputenc}
\usepackage{enumitem}
\usepackage{tikz}
\usetikzlibrary{positioning}
\usetikzlibrary{shapes.symbols,arrows.meta,positioning}

\definecolor{procFill}{RGB}{240,248,255}
\definecolor{procDraw}{RGB}{70,130,180}
\definecolor{procText}{RGB}{30,50,80}
\definecolor{fileFill}{RGB}{255,250,240}
\definecolor{fileDraw}{RGB}{210,180,140}
\definecolor{fileText}{RGB}{100,80,50}

\graphicspath{{original-thesis/figures/}}

\begin{document}

\title{Closing the Gap: Automated Discovery of Secure Dockerfile Reference Standards via Semantic Clustering in Enterprise Inner Source}

\author{Jessica Hösl}
\orcid{0009-0000-3000-9573}
\affiliation{%
  \institution{Technical University of Munich}
  \city{Munich}
  \country{Germany}
}
\email{jessica.hoesl@tum.de}

\author{Benedikt Hofmann}
\orcid{0000-0002-3905-5143}
\affiliation{%
  \institution{Siemens AG}
  \city{Munich}
  \country{Germany}
}
\email{hofmann.benedikt@siemens.com}

\author{Patrick Stöckle}
\correspondingauthor
\orcid{0000-0003-0193-5871}
\affiliation{%
  \institution{Siemens AG}
  \city{Munich}
  \country{Germany}
}
\email{patrick.stoeckle@siemens.com}

\renewcommand{\shorttitle}{Closing the Gap: Automated Discovery of Secure Dockerfile Reference Standards via Semantic Clustering ...}

\begin{abstract}
Containerization dominates enterprise software delivery,
yet Dockerfiles that assemble container images frequently harbor security
misconfigurations and structural technical debt.
This problem is poorly understood in corporate inner-source environments,
where proprietary context and isolated governance prevent direct application
of open-source findings.

We present an automated, six-stage pipeline that:
(1) crawls an enterprise GitLab instance,
(2) enriches each Dockerfile with static security and quality metrics
(Hadolint, ShellCheck, Trivy) and lifecycle data,
(3) groups functionally identical workloads using LLM-generated semantic
descriptions and HDBSCAN,
and (4) quantifies the optimization gap against cluster-internal reference
implementations.

Applied to 11,470 Dockerfiles from over 6,200 repositories at a single large
industrial company, we find a systemic deficit:
99\% of files contain at least one security misconfiguration,
80.8\% violate Dockerfile best practices,
and the median artifact has not been revised for 838~days.
Despite this, high-quality reference implementations already exist within
83\% of functional clusters.
Adopting these internal standards would increase the average security posture
score by 60.4\% without developing any new templates.
These findings, grounded in one organization's inner-source ecosystem, provide
a data-driven foundation for future automated, context-aware recommender
systems targeting enterprise supply-chain security; whether the observed
technical-debt distribution and optimization gap generalize to other
enterprises remains an open question for future multi-organization study.
\end{abstract}



\keywords{Dockerfile security, technical debt, inner-source, mining software
  repositories, LLM, semantic clustering, container security, DevSecOps}

\maketitle

\section{Introduction}
\label{sec:intro}

Containerization has become the cornerstone of modern software delivery,
with industry surveys reporting over 90\% adoption in production
environments~\cite{cncf_survey_2024}.
Dockerfiles---the configuration files that assemble container images---serve
as executable blueprints for production workloads, CI/CD pipelines,
and development environments alike.
Writing secure, maintainable Dockerfiles is non-trivial:
the format conflates system administration, shell scripting,
and dependency management into a single artifact,
demanding expertise that application developers often
lack~\cite{rosa_fixing_2024}.
The result is a documented accumulation of structural \emph{smells},
security misconfigurations, and lifecycle neglect in the broader
ecosystem~\cite{cito_empirical_2017,wu_characterizing_2020,haque_well_2022}.

\textbf{The enterprise inner-source gap.}
Virtually all large-scale empirical container research targets public GitHub
or Docker Hub data~\cite{cito_empirical_2017,eng_revisiting_2021,lin_large-scale_2020,wu_characterizing_2020}.
Enterprise inner-source environments---where thousands of teams share code via
an internal code hosting platform (e.g., GitLab, Bitbucket) under proprietary guidelines---
present a fundamentally different context.
Internal images often embed company-specific hardened base images,
carry organizational naming conventions,
and lack the community governance present in popular open-source projects.
Consequently, the volume, distribution, and remediability of security debt
within a large corporation's container ecosystem has not been systematically studied.

\textbf{Problem.}
The absence of a methodology to group enterprise Dockerfiles by
\emph{functional intent} prevents meaningful cross-team benchmarking.
Without such grouping, it is impossible to identify which high-scoring internal
configurations could serve as reference standards for their lower-quality
functional peers.
AI-driven remediation systems~\cite{ksontini_refactoring_2025,ye_llmsecconfig_2025, shabani2025flakiness}
are the most promising vehicles for scaled optimization,
relying on techniques like In-Context Learning (ICL) and Retrieval-Augmented Generation (RAG) to automate repairs.
However, to successfully generate context-aware code,
these models require high-quality, domain-specific reference demonstrations,
which do not exist in public datasets and are currently buried in organizational silos.

\textbf{Contributions.}
This paper presents an automated six-stage pipeline deployed in the
context of an internal GitLab instance of a major industrial company.

\begin{enumerate}[leftmargin=*, topsep=2pt, itemsep=1pt]
  \item \textbf{Semantic clustering methodology.}
    Enterprise Dockerfiles grouped by workload type
    based on LLM-generated functional descriptions of their intent,
    decoupling intent from syntactic implementation.
  \item \textbf{Composite Security Posture Score.}
    Linter warnings, security scans, base image footprint,
    and lifecycle metrics are unified into a normalized
    $[0,1]$ ranking that quantifies the security and maintainability.
  \item \textbf{First enterprise-scale empirical characterization.}
    11,470~Dockerfiles across 251~functional clusters,
    with a quantified optimization gap of 60.4\% achievable
    from existing internal data.
\end{enumerate}

\section{Background and Related Work}
\label{sec:related}

\subsection{Dockerfile Quality and Security Debt}

An extensive body of mining software repository research has characterized the open-source Dockerfile
landscape.
\citeauthor{cito_empirical_2017} identified missing version pinning as the dominant smell
in 70,000+ GitHub Dockerfiles~\cite{cito_empirical_2017};
\citeauthor{wu_characterizing_2020} confirmed widespread smell co-occurrence in 6,334
projects~\cite{wu_characterizing_2020};
\citeauthor{eng_revisiting_2021} revisited these findings with 9.4M Dockerfiles across seven years,
observing slow but measurable quality improvement~\cite{eng_revisiting_2021}.

On the security side,
\citeauthor{haque_well_2022} found that 91.6\% of open-source projects
inherit vulnerabilities from base images~\cite{haque_well_2022},
and \citeauthor{shi_dr_2025} recently extended this to 93.7\% of influential Docker Hub
images~\cite{shi_dr_2025}.
\citeauthor{opdebeeck_docker_2023} showed that 70\% of child images use a parent
more than five months out of date~\cite{opdebeeck_docker_2023}.
\citeauthor{rosa_mining_2025} established a taxonomy of 46 optimization strategies
by mining 1,026 improvement commits~\cite{rosa_mining_2025},
while \citeauthor{ksontini_refactorings_2021} catalogued Docker-specific technical debt types
from 68 open-source projects~\cite{ksontini_refactorings_2021}.

However, none of these studies target corporate inner-source environments, where different
base image profiles, organizational standards and requirements (e.g., license compliance),
and isolated development silos may lead to a distinct distribution of smells, misconfigurations,
and maintenance patterns.
Importantly, corporate environments also alter how the findings of standard tools such as
Hadolint and Trivy should be interpreted.
Company-hardened base images may pre-address certain Trivy security checks at the registry
level---for example, a proprietary base image may already enforce a non-root user---while
introducing internal security policies that tool defaults do not recognize.
Internal package mirrors can render Hadolint version-pinning warnings (e.g., \texttt{DL3008})
misleading when internal mirrors guarantee a fixed dependency resolution.
Organizational naming conventions cause Trivy and Hadolint to misclassify base images that
use private registry prefixes, inflating false-positive counts.
These factors collectively mean that open-source findings cannot be directly extrapolated
to the corporate setting: an empirical study targeting a corporate inner-source instance is
necessary to establish a valid baseline.

\subsection{Automated Repair and LLM-Based Approaches}

Rule-based tools such as DockerCleaner~\cite{bui_dockercleaner_2023}
and Parfum~\cite{durieux_empirical_2024} automate correction of specific smells
but cannot resolve complex, context-dependent misconfigurations.
LLM-based approaches address this gap:
\citeauthor{ksontini_refactoring_2025} use In-Context Learning (ICL) with GPT-4o,
achieving 32\% median image-size reduction~\cite{ksontini_refactoring_2025},
while LLMSecConfig~\cite{ye_llmsecconfig_2025} repairs Kubernetes
misconfigurations with 94\% success.
\citeauthor{rosa_automatically_2023} trained a T5 model on 670,000 Dockerfiles for full generation
from requirements, but found it struggles with complex enterprise
configurations~\cite{rosa_automatically_2023}.

A critical limiting factor for ICL systems is retrieval corpus quality.
\citeauthor{ksontini_refactoring_2025} use BM-25 keyword matching~\cite{ksontini_refactoring_2025},
which cannot recognize functionally equivalent but syntactically diverse
configurations.
\citeauthor{zhang_recommending_2022} showed that AST-based embeddings improve base-image recommendation
but remain tied to structural syntax~\cite{zhang_recommending_2022}.

Enterprise codebases are substantially more heterogeneous than open-source repositories:
different teams independently adopt different base images, package managers,
and build patterns for the same workload type, making syntactic similarity a poor
proxy for functional equivalence.
High-Level Specification (HLS) grouping by base image and installed packages
partially addresses this but remains anchored to implementation artifacts:
it conflates functionally distinct workloads that share a common base image
(e.g., a Python web API and a Python data-science notebook both built on
\texttt{python:3.11}),
while splitting functionally identical containers that happen to use different
distributions (e.g., Alpine vs.~Debian for the same REST service).
Applying Hadolint and Trivy directly to the full corpus yields ecosystem-wide
statistics (RQ1) but cannot identify \emph{which} specific Dockerfile should
serve as a reference standard for \emph{which} other Dockerfile---the
precise mapping that automated remediation and knowledge-transfer tools require.
Semantic clustering provides this mapping by grouping on functional intent
rather than implementation.

\subsection{Relationship to Commercial Container Security Tooling}

Commercial platforms such as Snyk Container, Aqua Security (the vendor behind
Trivy, which we use directly in Stage~3), and JFrog Xray already provide
enterprise-scale vulnerability scanning and remediation suggestions for
container images.
These tools excel at per-image, per-layer vulnerability detection and
typically propose a single upstream fix (e.g., bump a base-image tag or a
package version) grounded in CVE databases and vendor advisories.
They operate, however, at the level of an individual image or dependency
graph: they do not group functionally equivalent Dockerfiles across an
organization's codebase, and they have no notion of an internally vetted
\emph{golden reference} that already satisfies a company's own hardening
conventions, licensing constraints, and base-image standards.
We view our contribution as complementary to, rather than competing with,
this class of tools: their scan output (Trivy, in our pipeline) is a direct
input to the Security Posture Score, while semantic clustering and
golden-reference discovery add the missing organizational layer---identifying
\emph{which} already-compliant internal configuration should be propagated to
\emph{which} functionally equivalent peer, a mapping that commercial
per-image scanners do not produce.

\subsection{Research Gap}

Three gaps remain:
(1)~\emph{contextual}---no large-scale study targets corporate inner-source;
(2)~\emph{methodological}---as discussed above, recommendation frameworks lack semantic awareness
for functionally heterogeneous enterprise configurations,
requiring grouping by intent rather than syntax to map specific remediation targets;
(3)~\emph{foundational}---no approach systematically extracts high-quality
internal reference implementations to seed automated remediation.
Our work addresses all three.

\section{Industrial Context and Problem}
\label{sec:context}

The study was conducted at a global industrial technology company
with a self-hosted GitLab instance.
Development is organized across autonomous departments that can share code
via inner-source\cite{10.5555/3290281.3290310,10.1145/3611648}.
We only analyzed inner source repositories---those accessible to all employees---to capture the shared ecosystem
where cross-team knowledge transfer is possible, which included approximately 44,000 repositories.
Teams create their Dockerfiles or fork and extend base configurations but rarely re-synchronize with
functional peers.
This model creates \emph{silos of quality}, i.e., a highly optimized, company-hardened .NET container
built by a security-focused team coexists with dozens of misconfigured functional equivalents written by
teams lacking that expertise, although the others could access and adopt the high-quality reference
if they knew it existed.

Three concrete operational costs motivate automated remediation:

\textbf{Attack surface.}
Each container running as root or inheriting an unpatched base image
is a potential lateral-movement entry point.

\textbf{Maintenance burden.}
With dormant or abandoned Dockerfiles, engineering teams face growing hidden liability
as upstream packages accumulate CVEs while configurations remain static.

\textbf{Duplication of effort.}
Functional clusters contain dozens to hundreds of independent reimplementations
of the same workload type.
Systematic standardization can eliminate this redundancy,
reducing the number of unique artifact configurations requiring audit and
maintenance.

Prior to this work, no tooling existed to quantify this debt, map its
distribution, or extract remediation targets from existing internal repositories.

\subsection*{Research Questions}

This study is structured around three operationally motivated research questions:

\begin{description}[leftmargin=2em, topsep=2pt, itemsep=3pt, style=nextline]
  \item[\textbf{RQ1 --- Ecosystem State.}]
    \emph{What is the current security posture of Dockerfiles in a large
    corporate inner-source environment, as characterized by static analysis
    metrics and lifecycle data?}
    We establish the first empirical baseline for a corporate container
    ecosystem,
    quantifying smell prevalence, misconfiguration rates, and maintenance
    neglect.
  \item[\textbf{RQ2 --- Semantic Clustering Quality.}]
    \emph{Can LLM-generated functional descriptions, combined with
    density-based clustering, effectively group enterprise Dockerfiles by
    workload type?}
    We evaluate whether semantic grouping produces tighter functional cohesion
    than syntactic baselines---a prerequisite for valid intra-cluster
    benchmarking.
  \item[\textbf{RQ3 --- Addressable Remediation Potential.}]
    \emph{How much of the measured security debt can be resolved through
    internal knowledge transfer, i.e., by promoting configurations that already
    exist within the same functional cluster?}
    We quantify the per-file and ecosystem-wide optimization gap,
    and identify the high-priority clusters where automated remediation can
    have the largest measurable impact.
\end{description}

RQ1 is addressed in Section~\ref{sec:results:smells},
RQ2~in Section~\ref{sec:results:clustering},
and RQ3~in Section~\ref{sec:results:debt}.

\section{Approach}
\label{sec:approach}

\begin{figure}[t]
  \centering
  \resizebox{\columnwidth}{!}{%
  \begin{tikzpicture}[
      node distance=0.5cm and 2.1cm,
      process/.style={rectangle, draw=procDraw, fill=procFill, text=procText,
        rounded corners=3pt, minimum width=3.8cm, minimum height=0.95cm,
        align=center, font=\sffamily\bfseries\scriptsize, line width=0.7pt},
      artifact/.style={shape=tape, tape bend top=none, tape bend bottom=out,
        draw=fileDraw, fill=fileFill, text=fileText,
        minimum width=2.5cm, inner sep=3pt, align=center,
        font=\tiny\ttfamily, line width=0.5pt},
      arrow/.style={->, color=procDraw, line width=0.9pt,
        >={Stealth[round, length=2.5mm, width=1.8mm]}, rounded corners=4pt}
  ]
    \node[process] (step1) {1. Data Acquisition\\[-0.05em]{\normalfont\tiny (GitLab API Crawler)}};
    \node[artifact, below=of step1] (file1) {raw\_data.jsonl};
    \node[process, below=of file1] (step2) {2. Preprocessing\\[-0.05em]{\normalfont\tiny (Validation, Parsing, Normalization)}};
    \node[artifact, below=of step2] (file2) {clean\_data.parquet};
    \node[process, below=of file2] (step3) {3. Feature Extraction\\[-0.05em]{\normalfont\tiny (Quality \& Security Metrics)}};
    \node[artifact, below=of step3] (file3) {enriched.parquet};
    \node[process, right=of step1] (step4) {4. Semantic Analysis\\[-0.05em]{\normalfont\tiny (LLM Descriptions + Clustering)}};
    \node[artifact, below=of step4] (file4) {clustered.parquet};
    \node[process, below=of file4] (step5) {5. Golden Selection\\[-0.05em]{\normalfont\tiny (Reference Identification)}};
    \node[artifact, below=of step5] (file5) {ranked.parquet};
    \node[process, below=of file5] (step6) {6. Ecosystem Evaluation\\[-0.05em]{\normalfont\tiny (Optimization Gap Analysis)}};
    \draw[arrow] (step1) -- (file1);
    \draw[arrow] (file1) -- (step2);
    \draw[arrow] (step2) -- (file2);
    \draw[arrow] (file2) -- (step3);
    \draw[arrow] (step3) -- (file3);
    \draw[arrow] (step4) -- (file4);
    \draw[arrow] (file4) -- (step5);
    \draw[arrow] (step5) -- (file5);
    \draw[arrow] (file5) -- (step6);
    \draw[arrow] (file3.east) -- ++(1.9,0) |- (step4.west);
  \end{tikzpicture}%
  }
  \caption{Six-stage pipeline with intermediate data artifacts.
    Each stage persists results as Parquet snapshots for reproducibility.}
  \Description{Two-column diagram of a six-stage data pipeline. The left column shows stages 1--3 (Data Acquisition, Preprocessing, Feature Extraction) connected by arrows and data file artifacts. The right column shows stages 4--6 (Semantic Analysis, Golden Selection, Ecosystem Evaluation). An arrow connects the enriched artifact from stage 3 into stage 4 on the right.}
  \label{fig:pipeline}
\end{figure}
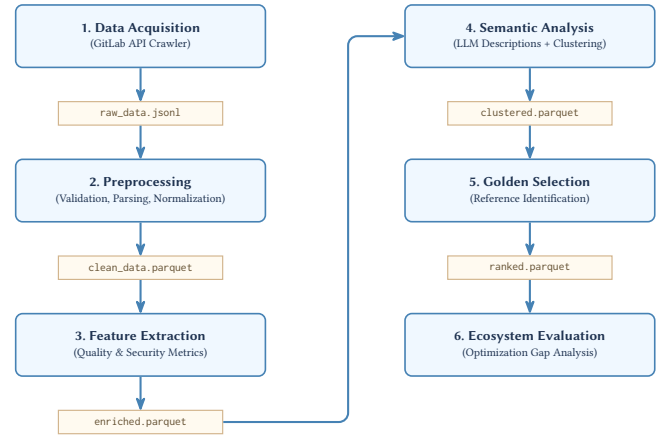

The pipeline consists of six sequential stages implemented in Python,
with specialized tool invocations (Go-based Moby parser, Hadolint, ShellCheck,
Trivy, crane) orchestrated centrally.
Intermediate results are persisted as Apache Parquet snapshots
for independent validation and reproducibility.
Figure~\ref{fig:pipeline} illustrates the complete pipeline.

\subsection{Stage 1: Data Acquisition}

We collect the Dockerfiles using a custom crawler that interacts with the GitLab v4 API,
targeting the default branch of each project.
Named Dockerfile variants (e.g., \texttt{Dockerfile.base}) are included;
template files (Jinja2 placeholders), documentation,
and Windows-image Dockerfiles are excluded.
The crawler collects:
project metadata, CI/CD pipeline configuration,
and container registry information.

Furthermore, Git commit history is extracted via partial cloning to capture file-specific revision timelines
without downloading blob content to optimize bandwidth and storage.
The resulting raw dataset comprised \textbf{12,002} unique Dockerfiles
from 6,247 repositories.

\subsection{Stage 2: Preprocessing and Normalization}

Raw records undergo integrity checks
(empty, malformed, invalid-syntax entries removed),
yielding \textbf{11,470} cleaned Dockerfiles.
Each file is parsed into an AST via a custom Go executable wrapping
the official \texttt{moby/buildkit}\footnote{\href{https://github.com/moby/buildkit}{github.com/moby/buildkit}} parser,
ensuring syntax parity with the Docker build engine
for heredoc and multi-stage syntax.

Build-time \texttt{ARG}/\texttt{ENV} variables are resolved from
associated CI/CD configurations,
reducing unresolved base-image variables by 20\%.
Normalized AST strings---comments stripped, multi-line \texttt{RUN} blocks
concatenated---are stored for downstream analysis.

\subsection{Stage 3: Feature Extraction and Scoring}
\label{sec:approach:scoring}

We compute six normalized metrics spanning three dimensions:

\textbf{Static Code Quality.}
Hadolint\footnote{\href{https://github.com/hadolint/hadolint}{github.com/hadolint/hadolint}} identifies Dockerfile-specific smells (DL-series rules).
ShellCheck\footnote{\href{https://www.shellcheck.net/}{shellcheck.net}} analyzes concatenated \texttt{RUN} blocks,
detecting cross-instruction shell errors that Hadolint misses in isolation.
Both are normalized as \emph{defect density} (violations per instruction),
clipped at population 95th-percentile saturation points
($D_\mathrm{max}^\mathrm{HL}\!=\!1.11$, $D_\mathrm{max}^\mathrm{SC}\!=\!0.25$)
to prevent extreme outliers from compressing the score of the addressable majority:
\[
  N_\mathrm{HL}(d) = 1 - \min\!\left(\frac{v_\mathrm{HL}(d)/n(d)}{D_\mathrm{max}^\mathrm{HL}},\, 1\right)
\]
where $v_\mathrm{HL}(d)$ is the Hadolint violation count
and $n(d)$ the instruction count of Dockerfile $d$.
The ShellCheck component $N_\mathrm{SC}$ is computed analogously.
The main reason for using density rather than raw counts is to avoid penalizing longer
files that may naturally have more violations but are not necessarily of lower quality.

\textbf{Security Configuration.}
Trivy\footnote{\href{https://github.com/aquasecurity/trivy}{github.com/aquasecurity/trivy}} in configuration-scan mode identifies misconfigurations
(e.g., \texttt{DS002}: running as root, \texttt{DS004}: exposing privileged ports)
without building images.
Raw counts $v_\mathrm{TV}(d)$ are $\ln(x\!+\!1)$-transformed and saturated at 10 misconfigurations:
\[
  N_\mathrm{TV}(d) = 1 - \min\!\left(\frac{\ln(v_\mathrm{TV}(d)+1)}{\ln(11)},\, 1\right)
\]
Running Hadolint and Trivy in parallel is intentional:
the former flags stylistic violations (\emph{explicit} \texttt{USER root}),
the latter flags security state (\emph{implicit} root inherited from a base
image),
capturing both noisy and silent risks.

\textbf{Supply Chain and Lifecycle.}
Base image compressed size $s(d)$ (queried via crane\footnote{\href{https://github.com/google/go-containerregistry/tree/main/cmd/crane}{github.com/google/go-containerregistry}})
is log-normalized over a $[s_\mathrm{min} = 10\,\text{MB}, s_\mathrm{max} = 2\,\text{GB}]$ range:
\[
  N_\mathrm{size}(d) = 1 - \min\!\left(\frac{\ln(\max(s(d), s_\mathrm{min})) - \ln(s_\mathrm{min})}
    {\ln(s_\mathrm{max}) - \ln(s_\mathrm{min})}, 1\right)
\]
Maintenance metrics are calibrated to literature-derived normative thresholds:
\emph{update frequency} (average revisions per year, $F_\mathrm{max}\!=\!12$~rev/year~\cite{cito_empirical_2017}) and
\emph{recency} (days since last commit, $R_\mathrm{max}\!=\!730$~days~\cite{haque_well_2022}).
A higher revision frequency and lower recency (more recent)
each contribute positively to the score.

The \textbf{Security Posture Score} aggregates these six metrics:
\[
  S = \sum_{i=1}^{6} w_i \cdot N_i(m_i), \qquad S \in [0,1]
\]
with weights listed in Table~\ref{tab:weights}:
Trivy and Hadolint each receive 0.25 (security risk and code quality),
update frequency 0.20, base image size 0.15, recency 0.10,
and ShellCheck 0.05.
This configuration prioritizes immediately exploitable risks over stylistic
issues.

\begin{table}[t]
  \small\centering
  \caption{Security Posture Score: weights and normalization limits.}
  \label{tab:weights}
  {\setlength{\tabcolsep}{4pt}%
  \begin{tabular}{@{}lccl@{}}
    \toprule
    \textbf{Metric} & \textbf{$w_i$} & \textbf{Limit} & \textbf{Rationale} \\
    \midrule
    Trivy (security)   & 0.25 & $>\!10$ misconfigs  & Immediate exploitability risk \\
    Hadolint (quality) & 0.25 & density $>\!1.11$   & Prerequisite for secure builds \\
    Update frequency   & 0.20 & $>\!12$/year        & Ability to respond to threats \\
    Base image size    & 0.15 & $>\!2$\,GB (log)    & Attack surface / bloat \\
    Recency            & 0.10 & $>\!730$ days       & Staleness / vulnerability lag \\
    ShellCheck         & 0.05 & density $>\!0.25$   & Shell-specific logic errors \\
    \bottomrule
  \end{tabular}}%
\end{table}

To prevent the systematic penalization of technology stacks with inherently larger footprints,
such as machine learning or robotics workloads,
we compute a \emph{Relative Security Posture Score} $S_\mathrm{rel}$ for intra-cluster comparison.
This metric normalizes base image size using each cluster's specific 5th--95th percentile bounds,
ensuring the resulting optimization gap reflects remediable configuration debt rather than architectural requirements.

\subsection{Stage 4: Semantic Clustering}
\label{sec:approach:clustering}

\textbf{Why semantic descriptions?}
Grouping Dockerfiles by syntactic features over-indexes on implementation
details.
For example, two Python REST APIs built on Alpine/pip and Debian/conda would be split into
separate clusters,
while syntactically similar but functionally different configurations are
merged.
Inspired by limitations identified in prior retrieval
work~\cite{zhang_recommending_2022, ksontini_refactoring_2025},
we embed natural-language \emph{functional descriptions} instead.

\textbf{LLM description generation.} Each normalized Dockerfile AST is submitted
to Qwen2.5-30B-Instruct with the following
constrained prompt template:

\begin{center}
\begin{tikzpicture}
\node [
    fill=gray!5,
    text width=\dimexpr\linewidth-6pt\relax,
    inner xsep=3pt,
    inner ysep=8pt,
    font=\small\itshape,
    align=flush left
] {
Generate a Semantic Capability Summary for this Dockerfile.\\
Constraints:\\
1. Describe the Application Workload (e.g., ’Python Data Science’, ’Node.js API’) and Key Frameworks.\\
2. Describe the Functional Intent (what the container does).\\
3. NEGATIVE CONSTRAINT: Do NOT mention the Base Image OS (e.g. Ubuntu, Alpine, Debian).\\
4. NEGATIVE CONSTRAINT: Do NOT mention specific version numbers.\\
5. NEGATIVE CONSTRAINT: Do NOT mention build-time cleanup or security flags.\\
6. Output exactly 2-3 concise sentences.
};
\end{tikzpicture}
\end{center}

The LLM is internally hosted to avoid data leakage and ensure compliance with the
company's data governance policies.
Temperature $T\!=\!0.1$ minimizes output variance across repeated runs.
Files exceeding 9,000 characters are truncated in the middle, preserving
the \texttt{FROM} header and \texttt{CMD}\slash\texttt{ENTRYPOINT} instructions
to retain workload-identification signal. The exclusion of OS distribution
and version numbers is deliberate: including them would allow the
embedding model to cluster by implementation artefacts rather than by
functional intent, defeating the semantic grouping objective.

\textbf{Embedding and clustering.}
Descriptions are embedded with \texttt{all-mpnet-base-v2}
(Sentence-Transformers, $\mathbb{R}^{768}$).
Dimensionality is reduced to 50 via UMAP (cosine metric, $k\!=\!15$,
$\text{min\_dist}\!=\!0$).
HDBSCAN (minimum cluster size~15, minimum samples~3) identifies dense
functional groups; sparse points are labelled as noise~\cite{campello2013density}.

We selected HDBSCAN over partitioning methods (e.g., k-means) and
agglomerative clustering for three reasons specific to an enterprise
Dockerfile corpus of unknown functional granularity.
First, the number of distinct workload types is not known a priori and
cannot be reliably estimated in advance;
HDBSCAN does not require the target cluster count $k$ to be fixed,
unlike k-means.
Second, functional cluster sizes are highly imbalanced
(from single-digit niche workloads to the 494-file ASP.NET cluster in
Table~\ref{tab:top10}); density-based clustering accommodates this
variable density better than centroid-based methods,
which implicitly assume roughly spherical, similarly sized clusters.
Third, LLM-generated descriptions for atypical or highly customized
Dockerfiles are occasionally ambiguous or idiosyncratic;
HDBSCAN's explicit noise label lets such points remain unclustered rather
than forcing them into an ill-fitting group, which would otherwise silently
corrupt intra-cluster comparisons.
We did not empirically ablate this choice against alternative clustering
algorithms on the full corpus;
Section~\ref{sec:threats} discusses this and related methodological
robustness questions as an explicit threat to validity.

\subsection{Stages 5--6: Golden Selection and Optimization Gap}
\label{sec:approach:gap}

Within each cluster $C_k$, the \emph{golden reference} $g_k$ is the Dockerfile
maximizing $S_\mathrm{rel}$. The \emph{Optimization Gap} for any file
$d_i \in C_k$ is:
\[
  \Delta_\mathrm{opt}(d_i) = S_\mathrm{rel}(g_k) - S_\mathrm{rel}(d_i),
  \quad g_k = \arg\max_{j \in C_k} S_\mathrm{rel}(d_j)
\]

A \emph{conservative P90 target} replaces $g_k$ with the 90th-percentile
intra-cluster score, accounting for configurations that score highly only through
extreme architectural constraints (e.g., \texttt{FROM scratch} static binaries)
that cannot be generalized across the full cluster.

\section{Evaluation}
\label{sec:results}

\subsection{Dataset Overview}

The final dataset contains \textbf{11,470} Dockerfiles from \textbf{6,247}
repositories.
The top 20 base images cover 67.3\% of all files; \texttt{python},
\texttt{node}, \texttt{alpine}, and \texttt{ubuntu} are the four most prevalent.
Base image sizes span from virtually empty (\texttt{scratch}) to 10.75~GB, with a
median of 47~MB and a 95th percentile of 471~MB.

The industrial base-image profile differs markedly from public-repository
studies~\cite{lin_large-scale_2020}: \texttt{aspnet} ranks first by cluster size
(ahead of \texttt{python} and \texttt{node}), reflecting the dominance of
enterprise web services, and two robotics/embedded clusters have no meaningful
counterparts in open-source datasets. Table~\ref{tab:top10} presents the ten
most populous functional clusters, spanning six distinct technology stacks that
collectively represent 17.5\% of the dataset.

\begin{table*}[t]
  \small\centering
  \caption{Top 10 functional clusters by file count. Dom.\ = dominant base image
    and its share within the cluster. Together these ten clusters cover 17.5\%
    of all Dockerfiles across six distinct technology stacks.}
  \label{tab:top10}
  \begin{tabular}{rlp{3.0cm}p{8.5cm}}
    \toprule
    \textbf{ID} & \textbf{$N$} & \textbf{Dom.\ Base Image} & \textbf{Functional Domain} \\
    \midrule
    27  & 494 & \texttt{aspnet} (61\%)  & ASP.NET Core web applications and REST API services \\
    215 & 262 & \texttt{python} (24\%)  & ML model development and inference (TensorFlow/PyTorch) \\
    191 & 196 & \texttt{ubuntu} (26\%)  & C/C++ compilation and build toolchains \\
    169 & 172 & \texttt{debian} (40\%)  & Embedded systems C/C++ cross-compilation (ARM) \\
    147 & 147 & \texttt{alpine} (16\%)  & Kubernetes operations and infrastructure automation (Helm) \\
    62  & 132 & \texttt{ros} (23\%)     & Real-time robotics and sensor processing (ROS~2) \\
    151 & 101 & \texttt{tomcat} (20\%)  & Identity, authentication, and secure token management \\
    210 &  86 & \texttt{alpine} (34\%)  & High-performance compiled Go microservices \\
    76  &  84 & \texttt{python} (17\%)  & Security scanning and automated vulnerability assessment \\
    143 &  82 & \texttt{nginx} (43\%)   & Frontend web applications and static asset delivery \\
    \bottomrule
  \end{tabular}
\end{table*}

\begin{figure}[htb]
  \centering
  \includegraphics[width=\columnwidth]{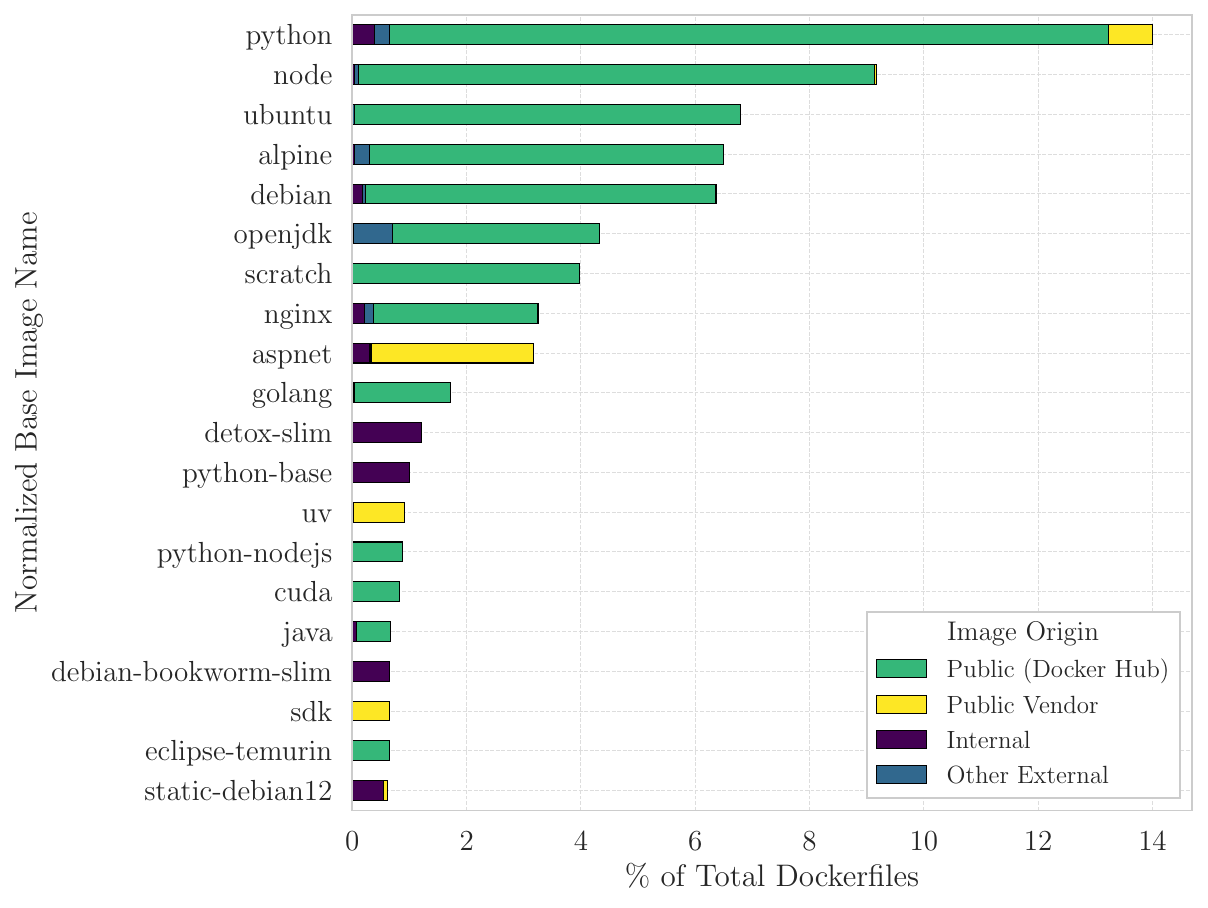}
  \Description{}
  \caption{Distribution of base image sizes across the dataset.}
  \label{fig:baseimg}
\end{figure}

Beyond lifecycle concerns, the ecosystem's architectural footprint reveals significant differences in base image usage, which directly affect operational efficiency and attack surface.
The top 20 base images account for 67.3\% of all Dockerfiles, with \texttt{python}, \texttt{node}, \texttt{alpine}, and \texttt{ubuntu} being the most prevalent (Figure \ref{fig:baseimg}).
Base image sizes exhibit a right-skewed distribution: the median is 47 MB (indicating a preference for optimized images), while the 95th percentile reaches 471 MB and the maximum extends to 10.75 GB (a machine learning \texttt{acpt-pytorch-cuda} image).
This underscores the need for context-specific optimization: 50 MB is substantial for \texttt{alpine}-based microservices but negligible for GPU workloads.

\subsection{Clustering Validation}
\label{sec:results:clustering}

Semantic clustering produced \textbf{251 clusters} covering 83\% of the dataset
(17\% noise, i.e., points not assigned to any cluster).
Table~\ref{tab:clustering} reports cluster-size weighted internal
cohesion metrics.

\begin{table}[t]
  \small\centering
  \caption{Clustering method comparison.
  Semantic (ours) is evaluated against syntactic baselines using Dockerfile Content (raw text)
  and HLS (High-Level Specification: base image + installed packages) as input representations.}
  \label{tab:clustering}
  \begin{tabular}{@{}lrrrr@{}}
    \toprule
    \textbf{Method} & \textbf{\#Clusters} & \textbf{Noise} &
      \textbf{Img Entropy} & \textbf{Pkg Overlap} \\
    \midrule
    Semantic (ours) & 251 & 0.17 & 1.65 & 0.41 \\
    Content (raw)   & 257 & 0.21 & 1.20 & 0.46 \\
    HLS (base+pkgs) & 309 & 0.14 & 0.79 & 0.59 \\
    \bottomrule
  \end{tabular}
\end{table}

Although the semantic approach yields higher base-image entropy than syntactic
baselines (1.65 vs.~ 0.79--1.20),
this is by design:
the model groups heterogeneous implementations of the same workload.
The global dataset entropy is 4.32;
semantic clustering reduces it by \textbf{62\%},
confirming meaningful functional cohesion.
The raw-content baseline produces 21\% noise,
indicating it over-fits to syntactic variation.

We emphasize that Table~\ref{tab:clustering} isolates the effect of the
\emph{input representation} (semantic descriptions vs.\ raw content vs.\ HLS)
while holding the clustering algorithm (HDBSCAN) fixed;
it is not a comparison across clustering algorithms.
Whether an alternative algorithm (e.g., k-means or agglomerative clustering)
applied to the same semantic embeddings would yield comparable or superior
cohesion remains untested on this corpus and is discussed further in
Section~\ref{sec:threats}.

\begin{figure}[htb]
  \centering
  \includegraphics[width=0.7\columnwidth]{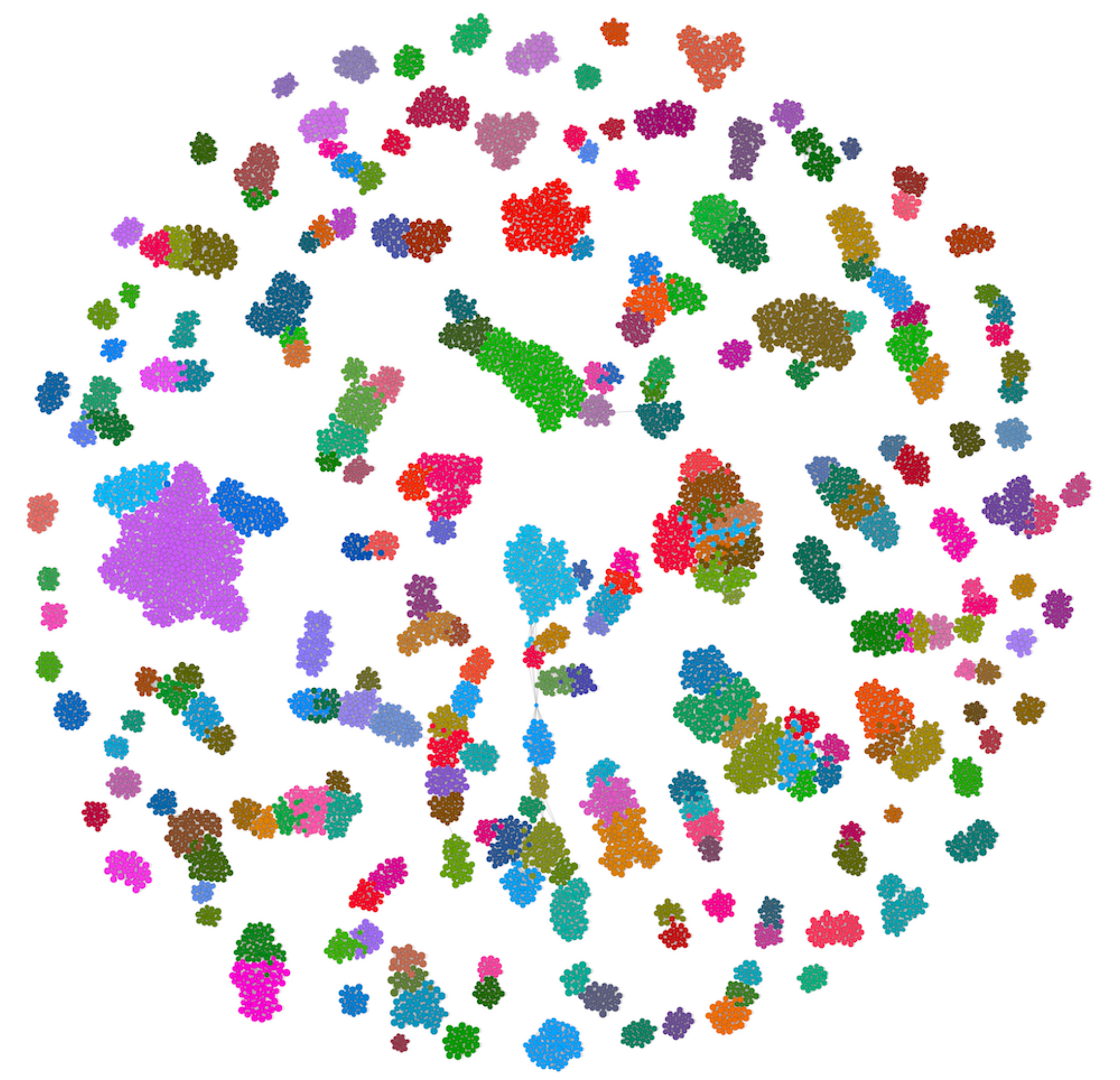}
  \Description{Force-directed graph showing 11,470 Dockerfiles as nodes, colored by Cluster ID. Distinct, color-coded islands represent functional clusters; edges connect files with embedding similarity above 0.6. The structure visually confirms semantic coherence and topological separation.}
  \caption{Force-Directed Graph of the Dockerfile Ecosystem, colored by Cluster ID.
    Nodes represent individual Dockerfiles; edges denote cosine embedding
    similarity $> 0.6$. Distinct topological clusters validate the semantic
    model's ability to isolate functionally coherent workloads.}
  \label{fig:graph:clusters}
\end{figure}

The force-directed graph visualization in Figure~\ref{fig:graph:clusters} provides both topological and evaluative
perspective on the clustered ecosystem.
Nodes represent individual Dockerfiles, and
edges denote cosine embedding similarities exceeding~0.6---a threshold chosen
to reduce visualization noise while preserving the distinct topological structure
of functional groups.

The resulting topology reveals tightly connected color-coded islands
(dense functional clusters), validating the algorithm’s ability to isolate semantically coherent workloads.
For example, the large violet cluster on the left (494 Dockerfiles, .NET applications)
correctly groups canonical \texttt{aspnet} images
and company-hardened enterprise derivatives
by their shared runtime purpose,
despite different registry origins and security scores (0.07--0.87).

Similarly, Node.js workloads are close together spatially
but subdivided into semantically distinct groups:
frontend frameworks (React, Next.js) are separated
from backend API gateways and build environments.
The clear, visually distinct grouping confirms that semantic descriptions
decouple functional intent from syntactic implementation.

\begin{figure}[htb]
  \centering
  \includegraphics[width=0.7\columnwidth]{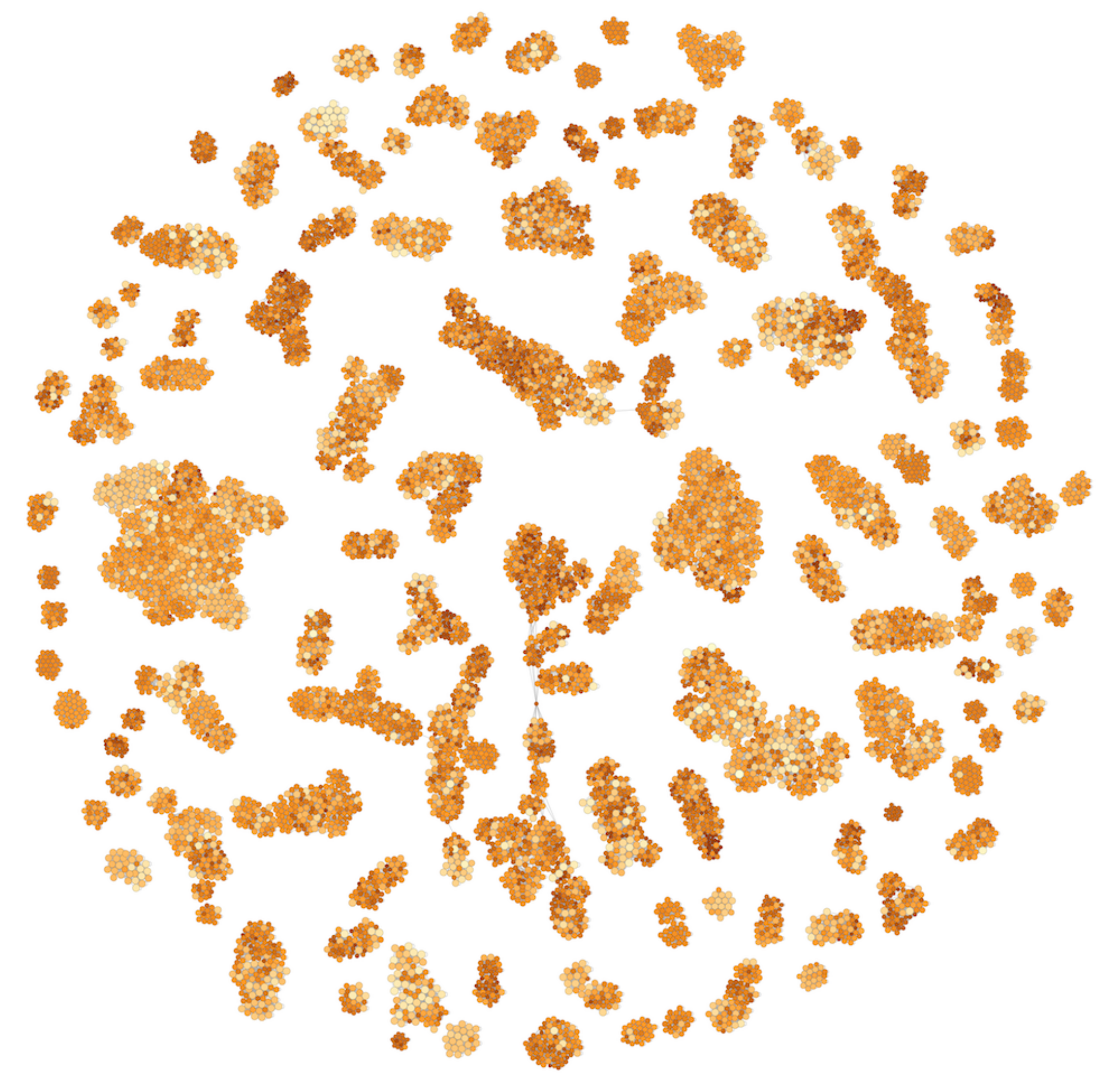}
  \Description{Force-directed graph showing 11,470 Dockerfiles as nodes, colored by relative security score from dark (low) to light (high). The same topological structure shows persistent within-cluster score heterogeneity.}
  \caption{Force-directed graph of the clustered Dockerfiles, colored by
    relative security posture score (dark~= low, light~= high). Within-cluster
    score spread is visible across virtually all functional groups, confirming
    that high-quality configurations rarely propagate to their lower-scoring
    cluster peers.}
  \label{fig:graph}
\end{figure}

Figure~\ref{fig:graph} shows the \emph{same} force-directed topology annotated by
relative score instead of cluster membership.
Each node remains a Dockerfile;
the topology is invariant.
The color gradient from dark (low~$S_\mathrm{rel}$) to light (high~$S_\mathrm{rel}$)
reveals the within-cluster quality variance:
in most clusters, a small subset of light nodes coexists with a large majority
of dark and medium nodes,
visually confirming the bi-modal quality distribution that the Optimization
Gap metric captures numerically.

\subsection{State of the Ecosystem}
\label{sec:results:smells}

\begin{figure}[htb]
  \centering
  \includegraphics[width=\columnwidth]{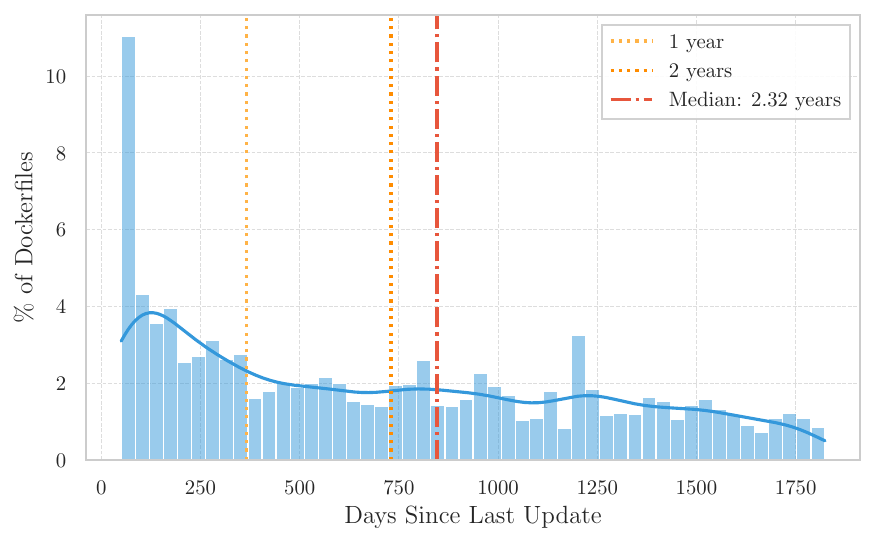}
  \Description{Histogram of days since last Git commit for 11,470 Dockerfiles. The distribution is right-skewed with a peak around 200--400 days; 70.2\% of files have not been updated in over a year.}
  \caption{Distribution of artifact recency (days since last update). 70.2\% of
    the ecosystem has not been updated in over a year.}
  \label{fig:recency}
\end{figure}

\begin{figure}[htb]
  \centering
  \includegraphics[width=\columnwidth]{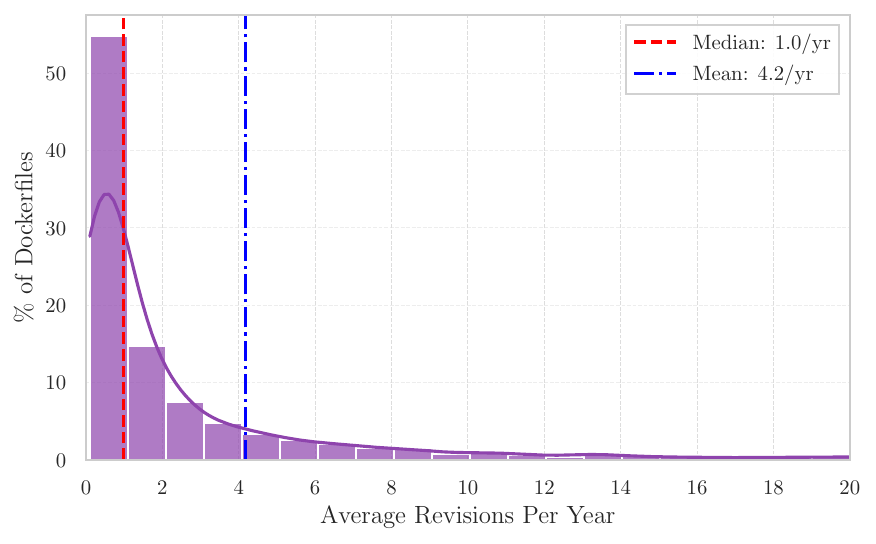}
  \Description{Histogram of Git revision counts per Dockerfile.}
  \caption{Distribution of Git revision counts per Dockerfile.}
  \label{fig:revisions}
\end{figure}

\textbf{Lifecycle neglect.}
Figure~\ref{fig:recency} illustrates the recency distribution
and Figure~\ref{fig:revisions} the revision-count distribution.
Both are strongly right-skewed.
While 30\% of files where updated within the last year,
the median Dockerfile has not been revised in \textbf{838~days} (2.3~years);
the oldest unmaintained file has not been updated for a decade.
While a lack of recent commits can indicate a stable,
feature-complete service rather than abandonment,
this extreme latency creates severe security risks.
In containerized environments, an unmaintained Dockerfile could pull outdated,
unpatched base OS layers and dependencies, leading to software decay,
even when the application code remains unchanged.
Applying a maintenance taxonomy:

\begin{itemize}[topsep=2pt, itemsep=0pt]
  \item \textbf{Dormant/Abandoned} (last commit $>$1~year): \textbf{70.2\%}
  \item \textbf{Actively Maintained} ($\geq 2$~rev/year, recent):
    \textbf{23.4\%}
  \item Stable or New: 6.4\%
\end{itemize}

Update frequency mirrors open-source findings~\cite{cito_empirical_2017}:
half the ecosystem is updated once or less per year,
yet the top 10\% exceed 9.8 revisions annually.
The bimodal character suggests that automated remediation would disproportionately
benefit the inactive majority.

\textbf{Smell prevalence.}
Hadolint identified \textbf{45,902 violations} across \textbf{80.8\%} of all
Dockerfiles (median: 2; maximum: 95).
While these violations are widespread, their distribution is highly concentrated.
The most common violations by prevalence are missing version pinning (\texttt{DL3008}, 29.4\% of
files),
consecutive \texttt{RUN} instructions (\texttt{DL3059}, 27.4\%),
and the absence of \mbox{\texttt{--no-install-recommends}} (\texttt{DL3015},
19.0\%).

\begin{figure}[htb]
  \centering
  \includegraphics[width=\columnwidth]{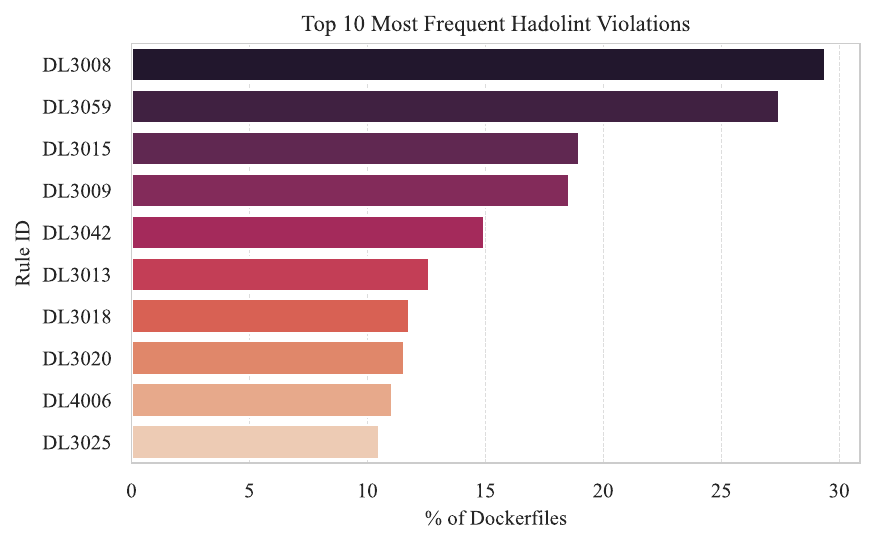}
  \Description{Horizontal bar chart showing the top 10 Hadolint rule violations by total count. DL3008 (apt-get without version pins) and DL3059 (redundant RUN instructions) are the most frequent, each exceeding 5,000 occurrences.}
  \caption{Top-10 Hadolint rule violations. Percentages reflect the prevalence of each rule across the total dataset ($N=11,470$).}
  \label{fig:hadolint}
\end{figure}

Figure~\ref{fig:hadolint} shows the top-10 violations by prevalence.
The data identifies two primary categories of technical debt:
Layer Optimization (e.g., \texttt{DL3059}, \texttt{DL3015}) and
Package-manager Pinning (\texttt{DL3008}, \texttt{DL3013}, \texttt{DL3016}).
Together, these categories affect over 50\% of the total files in the dataset.
While pinning rules are a critical component of supply-chain security,
their prevalence (appearing in 5,269 files) is matched by optimization smells (5,322 files).
This indicates that developers struggle equally with the security and the efficiency of their Dockerfiles.

\begin{figure}[htb]
  \centering
  \includegraphics[width=\columnwidth]{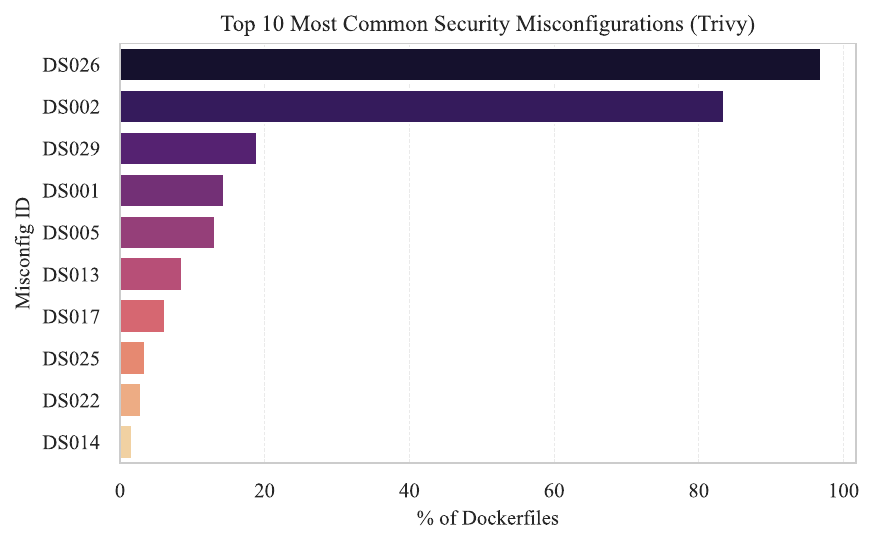}
  \Description{Horizontal bar chart of the top 10 Trivy misconfiguration findings. DS026 (missing HEALTHCHECK) and DS002 (running as root) are the two most frequent, each present in over 80\% of Dockerfiles.}
  \caption{Top-10 Trivy misconfigurations. \texttt{DS026} (missing HEALTHCHECK)
    and \texttt{DS002} (root user) co-occur in 81.7\% of files.}
  \label{fig:trivy}
\end{figure}

\textbf{Security misconfigurations.}
Trivy flagged \textbf{99\%} of Dockerfiles with at least one misconfiguration
(Figure~\ref{fig:trivy}).
The dominant pair---missing \texttt{HEALTHCHECK} (\texttt{DS026})
and running as root (\texttt{DS002})---appears simultaneously in \textbf{81.7\%}
of files,
producing a spike at exactly two misconfigurations (42.4\% of the dataset).
Running as root (\texttt{DS002}), present in 81.7\% of files,
is particularly concerning:
if a container process is compromised, an attacker would have root privileges within the container,
indicating that foundational security controls are systematically deferred to
downstream orchestration layers rather than embedded at the source.

\textbf{Composite score.}
The ecosystem achieves a mean $S_\mathrm{rel}$ of $\mathbf{0.48}$
($SD\!=\!0.16$, median~0.47)---centered near the midpoint of the $[0,1]$ scale.
The distribution is approximately normal with slight left skew:
approximately 6.3\% of files score above~0.75 (``acceptable security posture'')
and only 1.2\% exceed~0.85 (``strong posture''),
while 11.1\% score below~0.30.
This global symmetry masks important cluster-level heterogeneity:
the ML cluster (ID~215) has a mean of~0.40,
while the Go microservices cluster (ID~210) averages~0.56.
The weighted standard deviation within clusters is 0.13 on average,
confirming that meaningful intra-cluster spread---rather than across-cluster
divergence---is the primary driver of the optimization gap.

\subsection{Security Debt and Optimization Gap}
\label{sec:results:debt}

\textbf{Core result.}
For the 83\% of Dockerfiles successfully clustered,
the mean optimization gap is $\overline{\Delta}_\mathrm{opt}\!=\!0.29$.
The average clustered Dockerfile can increase its absolute score
by \textbf{29 percentage points}---a \textbf{60.4\% relative improvement}---simply
by adopting the best practices already present in its functional cluster.
The most severely under-performing 10\% face a gap exceeding 0.53:
alignment with the golden reference would more than double their security
posture.

Table~\ref{tab:optgap} compares optimization gaps across clustering methods.
Syntactic methods report a substantially lower gap (49.6\%/46.4\%) because they
trap insecure files with structurally similar but equally insecure neighbors,
hiding the existence of superior functional equivalents.
By grouping on intent rather than syntax,
semantic clustering compares configurations that serve the same purpose,
regardless of implementation differences.

\begin{table}[t]
  \small\centering
  \caption{Optimization gap by clustering method.}
  \label{tab:optgap}
  \begin{tabular}{@{}lrrr@{}}
    \toprule
    \textbf{Method} & \textbf{Total debt} & \textbf{Avg/file} &
      \textbf{Rel.\ improvement} \\
    \midrule
    Semantic (ours) & 2784.5 & 0.29 & 60.4\% \\
    Content (raw)   & 2212.4 & 0.24 & 49.6\% \\
    HLS             & 2244.3 & 0.23 & 46.4\% \\
    \bottomrule
  \end{tabular}
\end{table}

\textbf{Ecosystem-wide impact.}
If all clustered debt were resolved,
the global mean score (all $N\!=\!11,470$ files, including noise)
would rise from~0.48 to~\textbf{0.72}---a \textbf{50.6\%
global improvement}.
The conservative P90 target yields 34.5\% relative improvement for clustered
files and a \textbf{28.75\% global improvement}---still a substantial and
operationally realistic gain.

\begin{figure}[htb]
  \centering
  \includegraphics[width=\columnwidth]{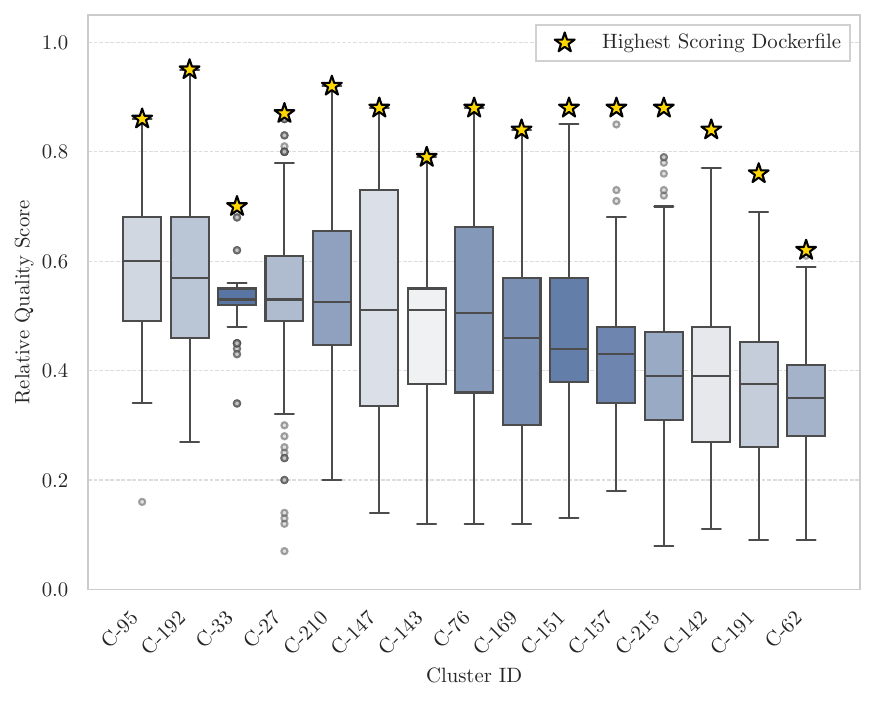}
  \Description{Box plots showing the distribution of relative security scores for the 15 largest clusters. Each box plot is annotated with a star marker indicating the golden reference (maximum score). A persistent gap between the interquartile range and the golden reference is visible in all clusters.}
  \caption{Security Posture Score distributions vs.\ maximum reference scores for
    the 15 largest clusters. Stars ($\star$) mark golden references. The
    persistent gap confirms systemic, not isolated, debt.}
  \label{fig:optgap}
\end{figure}

\textbf{Cluster-level variance.}
Figure~\ref{fig:optgap} visualizes the score distribution and golden reference
for the 15 largest clusters.
The persistent gap between each cluster's top quartile and its golden reference
($\star$) is visible across all workload types.
Within Kubernetes infrastructure automation (C-147),
security scores span the full range from near-zero to near-perfect,
while the dominant base image (\texttt{alpine}) accounts for only 16\% of the
cluster;
the remaining 84\% perform similar operational tasks on different
images.
This confirms that measured debt is not a property of a few outlier workloads
but is deeply embedded across all major categories.

\textbf{High-priority remediation clusters.}
Analysis of the bottom-15 clusters by mean score reveals two failure patterns.
\emph{Lifecycle abandonment}:
Python DevOps tooling (C-148) and JavaScript QA automation (C-103)
have median recencies of 1,384 and 1,246 days respectively.
While both exhibit extreme age, their technical debt profiles diverge:
C-148 contains a median of 61 Hadolint violations per file,
whereas C-103 reflects a lower but still elevated median of 9 violations.
\emph{Architectural sprawl}:
Robotics workloads (C-62) exhibit an extreme lack of standardization,
with base image footprints spanning three orders of
magnitude (26~MB to 8.2~GB),
indicating the absence of a unified engineering standard.
Flink/Kafka stream-processing (C-73) shows a
slightly more contained variance (28~MB to 535~MB).

Table~\ref{tab:highyield} presents the seven clusters with the highest cumulative
optimization gap that also contain an internal reference standard with
$S_\mathrm{max}\!\geq\!0.80$. These represent the highest-priority targets for
automated remediation: rich in debt, yet already possessing an good reference
configuration. The ASP.NET cluster (ID~27) alone concentrates 156.95 cumulative
gap points across 494~files; closing it costs zero new template development.
The ML inference cluster (ID~215) is the second-largest opportunity, with a
median score of only 0.39 despite the cluster's best configuration reaching
$S_\mathrm{max}\!=\!0.88$---a gap that reflects the widespread practice of
prototyping ML workloads without hardening before deployment.

\begin{table*}[t]
  \small\centering
  \caption{High-yield remediation targets: clusters with the highest cumulative
    optimization gap that already contain a viable internal reference
    ($S_\mathrm{max}\!\geq\!0.80$). $\sum\Delta_\mathrm{opt}$ is the total
    addressable security debt within the cluster.}
  \label{tab:highyield}
  \begin{tabular}{@{}rlp{6.8cm}ccc@{}}
    \toprule
    \textbf{ID} & \textbf{$N$} & \textbf{Functional Domain} &
      \textbf{$S_\mathrm{med}$} & \textbf{$S_\mathrm{max}$} &
      \textbf{$\sum\Delta_\mathrm{opt}$} \\
    \midrule
    27  & 494 & ASP.NET Core web applications and REST API services          & 0.53 & 0.87 & 156.95 \\
    215 & 262 & ML model development and inference (TensorFlow/PyTorch)      & 0.39 & 0.88 & 125.94 \\
    169 & 172 & Embedded systems C/C++ cross-compilation (ARM)               & 0.46 & 0.84 &  66.05 \\
    147 & 147 & Kubernetes operations and infrastructure automation (Helm)   & 0.51 & 0.88 &  52.12 \\
    151 & 101 & Identity, authentication, and policy enforcement (OAuth2)    & 0.44 & 0.88 &  42.89 \\
    157 &  81 & C/C++ development environments and Python build automation   & 0.43 & 0.88 &  36.38 \\
    142 &  74 & AWS infrastructure automation and cloud governance tooling   & 0.39 & 0.84 &  32.41 \\
    \bottomrule
  \end{tabular}
\end{table*}

\textbf{Clusters without viable references.}
The top 20 functional clusters have a median structural baseline
(median $S_\mathrm{max}$) of 0.90.
Conversely, the bottom 20 clusters currently
lack any reference implementation approaching a standard,
meaning no internal candidate is suitable for automated remediation. These
clusters consist largely of near-identical clones of a single low-quality
template, duplicated before a secure baseline was ever established. A targeted
investment in a single hardened base image per orphaned cluster would increase
the total addressable optimization potential by 7.7\% (215 cumulative debt
points across 422 files).

Table~\ref{tab:orphaned} identifies the five most important active clusters in
this category---active in the sense that recent commit history confirms ongoing
engineering investment. These are not abandoned workloads; they represent
live deployments lacking a quality anchor. The Internal Tool cluster (ID~12) is
particularly notable: with a median recency of 197~days, it is the most actively
maintained orphaned cluster, yet no member has reached $S_\mathrm{max}\!=\!0.75$.
This signals a systematic organizational gap---one that a purpose-built,
security-reviewed image could resolve for all current and future users.

\begin{table}[t]
  \small\centering
  \caption{Active clusters without a viable internal reference
    ($S_\mathrm{max}\!<\!0.75$). Despite recent development activity, no
    cluster member has reached sufficient quality to serve as a remediation
    template. Recency = median days since last commit.}
  \label{tab:orphaned}
  \begin{tabular}{@{}rp{2.2cm}rr r@{}}
    \toprule
    \textbf{ID} & \textbf{Functional Domain} & \textbf{$N$} &
      \textbf{$S_\mathrm{max}$} & \textbf{Recency} \\
    \midrule
    45  & Java Spring Boot (JAR)    & 39 & 0.56 & 245 d \\
    133 & NGINX Static Frontend     & 18 & 0.62 & 265 d \\
    214 & Node.js/TS Build \& Test  & 22 & 0.66 & 355 d \\
    12  & .NET/Node.js Internal Tool   & 29 & 0.67 & 197 d \\
    63  & Java JVM (container-opt.) & 28 & 0.67 & 245 d \\
    \bottomrule
  \end{tabular}
\end{table}

\section{Lessons Learned and Industrial Implications}
\label{sec:lessons}

\textbf{L1: Inner-source has structurally higher but unevenly distributed
maintenance activity.}
The top 10\% of inner-source Dockerfiles receive nearly 10 revisions per year
due to close coupling to production workloads.
Yet 70\% remain dormant.
The enterprise setting does not immunize organizations from the systemic
infrastructure neglect documented in public
repositories~\cite{cito_empirical_2017,eng_revisiting_2021}.

\textbf{L2: The problem is propagation, not expertise.}
The 60.4\% optimization gap does not reflect a lack of internal knowledge.
High-quality, hardened configurations already exist within 83\% of functional
clusters---built by security-aware teams.
The problem is that other teams writing functionally equivalent containers never
discover these standards.
This is exactly the retrieval problem that a semantics-aware recommender can
solve.

\textbf{L3: Semantic clustering yields a 14-percentage-point improvement over the
best syntactic baseline, and the LLM cost is a one-time batch investment.}
As seen in Table \ref{tab:optgap}, HLS-based clustering---the strongest syntactic baseline---reports a 46.4\% relative
optimization gap; raw-content clustering reports 49.6\%;
our semantic approach reports 60.4\%.
The 14-percentage-point improvement over HLS arises because semantic clustering
correctly bridges functionally equivalent containers that differ only in their
implementation choices (OS distribution, package manager),
exposing high-quality references that syntactic methods never associate with
their lower-quality functional peers.
The cost of this improvement is the one-time batch inference of LLM-generated
functional descriptions for the 11,470-file corpus using an internally hosted model
(Qwen2.5-30B-Instruct).
Because the model is self-hosted, there is no per-token billing;
the computation is a bounded, single-pass job that does not recur unless the corpus
is refreshed.
In contrast, the 14-percentage-point shortfall of HLS clustering is a \emph{permanent
structural deficit}: no amount of additional syntactic refinement can recover
functionally equivalent but syntactically dissimilar configurations.
For a 60\% improvement over an ~11,000-file corpus, we conclude that the
one-time LLM inference cost is justified; practitioners in resource-constrained
environments may nonetheless opt for HLS clustering as a lower-cost approximation
that still captures~46\% of the addressable debt.
Practitioners deploying ICL-based repair systems with syntactic
retrieval~\cite{ksontini_refactoring_2025} leave this meaningful improvement
unrealized by failing to bridge the gap between
distinct-but-equivalent configurations.

\textbf{L4: 83\% of the debt is addressable from existing data; semantic clustering
identifies 30\% more addressable debt per file than the syntactic baseline.}
Large organizations need not commission new secure templates to begin
remediation.
To quantify how much of this result depends on semantic clustering specifically:
under HLS-based clustering, 86\% of files are assigned to a cluster
(vs.~83\% for semantic)
but the mean addressable gap per file is only 0.23 (vs.~0.29 for semantic)---a
20.7\% reduction in identified optimization potential per artifact.
In absolute terms, HLS clustering surfaces 2,244 cumulative debt points
against semantic clustering's 2,785---a difference of 541 debt points that
would remain hidden without functional grouping.
The additional coverage comes from configurations that HLS incorrectly
separates or merges, causing their actual best functional-peer reference
to be invisible to the analysis.
Existing internal best practices are thus sufficient to address the overwhelming
majority of identified debt under either method;
the semantic approach ensures that more of that debt is correctly attributed
and that a richer set of valid reference implementations is surfaced.
Only the minority of orphaned clusters require targeted engineering
investment---and this analysis has identified precisely those clusters.

\textbf{Enabling downstream automation.}
The enriched, clustered dataset is a direct input to LLM-based repair
systems~\cite{ksontini_refactoring_2025,ye_llmsecconfig_2025}:
golden references serve as in-context demonstrations for enterprise-native
repair suggestions;
semantic cluster labels enable workload-aware retrieval,
directly addressing the BM-25 limitation;
and the Optimization Gap provides a prioritized list of files warranting
automated intervention.

\section{Threats to Validity}
\label{sec:threats}

\textbf{Construct.}
The Security Posture Score is a proxy for configuration hygiene,
not a runtime CVE count.
Equal weighting of all warnings regardless of severity may over-penalize
low-severity issues.
Weights and thresholds reflect expert judgment and are subject to sensitivity
analysis.
They were calibrated to align with industry best practices and internal security
standards, but alternative configurations could yield different absolute scores.

\textbf{Internal.}
Variable resolution reduced unresolvable base-image definitions to 5.1\%;
these files were retained to avoid bias against complex, parametrized
Dockerfiles.
The Git \texttt{--follow} flag was omitted,
potentially affecting recency for renamed files.
Hadolint's stricter parser may have discarded a small number of buildable but
non-standard Dockerfiles.

\textbf{External.}
The dataset originates from a single corporate environment.
Technical debt distribution, workload mix, and governance posture may differ
across organizations.
The study is a static snapshot;
longitudinal analysis of debt accumulation velocity remains future work.

\textbf{Clustering.}
No labeled ground truth is available for unsupervised clustering on proprietary
data.
We validated relative cohesion via base-image entropy and package Jaccard
similarity against syntactic baselines.
UMAP and HDBSCAN hyperparameters were tuned heuristically;
different settings alter cluster granularity and the measured Optimization Gap.
Our clustering-algorithm ablation is limited to input representation
(Table~\ref{tab:clustering}): we did not compare HDBSCAN against alternative
clustering algorithms (e.g., k-means, agglomerative, or spectral clustering)
applied to the same semantic embeddings, so we cannot rule out that a
different algorithm would produce different---possibly tighter or looser---functional
groupings and a correspondingly different Optimization Gap.
Relatedly, the reported results depend on two fixed model choices:
the sentence embedding model (\texttt{all-mpnet-base-v2}) used to vectorize
LLM-generated descriptions, and the LLM used to generate those descriptions
(Qwen2.5-30B-Instruct). We have not verified whether clustering outcomes are
stable under alternative embedding models or alternative LLMs; different
models could phrase functional descriptions with different granularity or
emphasis, which could shift cluster boundaries and noise rates. We treat
this as an open robustness question rather than an assumption, and outline
concrete ablation plans in Section~\ref{sec:future_work}.

\section{Future Work}
\label{sec:future_work}

We plan to use the enriched dataset resulting from this analysis as follows:

\textbf{LLM-based automated refactoring (ICL).}
The golden reference Dockerfiles are high-quality, company-vetted exemplars
in the exact technology stack of the target file.
Providing them as few-shot in-context examples to an LLM-based repair
system~\cite{ksontini_refactoring_2025} eliminates the need for public-corpus
retrieval and replaces it with semantically verified internal examples---directly
addressing the context gap identified
in~\cite{ye_llmsecconfig_2025}.
Such a repair system could be integrated into CI pipelines or applied to all files with
an optimization gap above a certain threshold as a batch remediation effort via pull requests.

\textbf{Missing Golden Reference Resolution.}
Some clusters lack a viable internal reference ($S_\mathrm{max}\!<\!0.75$).
For these, we create a secure baseline configuration through manual engineering effort,
guided by the cluster's functional domain and common base images.
This new baseline then serves as the golden reference for all cluster members,
enabling automated remediation for the 17\% of clusters currently without a viable internal exemplar.

\textbf{Longitudinal Debt Evolution and Operational Integration.}
The results in this paper are a static snapshot of a live, continuously
evolving ecosystem: new Dockerfiles are added, existing ones are edited, and
the company's hardened base images are periodically updated.
Future work will operationalize the pipeline as a recurring process rather
than a one-time study, along three complementary directions:
(1)~\emph{periodic full re-clustering}, re-running Stages~4--6 on a fixed
cadence (e.g., quarterly) to capture ecosystem-wide drift in workload
composition and golden-reference quality;
(2)~\emph{incremental assignment}, embedding and scoring newly added or
modified Dockerfiles against existing cluster centroids between full
re-clustering runs, flagging files that fit no known cluster as candidates for
manual triage rather than forcing a match; and
(3)~\emph{drift alerting}, monitoring when a cluster's current golden
reference is superseded (e.g., because the company's hardened base image
changes) or when a cluster's mean Optimization Gap increases, triggering
re-evaluation of that cluster's reference.
Together these mechanisms would reveal the velocity of debt accumulation, the
impact of remediation efforts, and the lifecycle dynamics of infrastructure as
code in an enterprise context, while keeping the golden-reference set current
as the ecosystem evolves.

\textbf{Deployment and Adoption Plan.}
Translating the 60.4\% optimization gap into realized security improvement
requires an adoption process, not merely the identification of golden
references.
We plan to prioritize outreach using the same ranking that drives
Table~\ref{tab:highyield}: clusters with the largest cumulative optimization
gap and a viable internal reference are contacted first, since they offer the
highest expected security return for the lowest engineering effort.
For each prioritized cluster, we plan to share the golden reference and a
per-file diff against it with the owning team through existing inner-source
communication channels (e.g., merge requests against the team's own
repository), rather than mandating adoption centrally.
Clusters without a viable reference (Table~\ref{tab:orphaned}) are handled
separately, as targeted engineering investment rather than propagation.
A follow-up measurement---re-running the Stage~3 scoring on the same
repositories after an adoption window---would let us report actual, rather
than theoretical, improvement, complementing the longitudinal re-analysis
described above.

\textbf{Clustering and Model Robustness Ablations.}
To close the methodological gaps identified in Section~\ref{sec:threats},
we plan a systematic robustness study along three axes:
(1)~\emph{clustering algorithm}---re-clustering the same semantic embeddings
with k-means, agglomerative, and spectral clustering, comparing cluster
count, noise rate, base-image entropy, and package overlap against the
HDBSCAN results reported in Table~\ref{tab:clustering};
(2)~\emph{embedding model}---repeating the pipeline with alternative
sentence-embedding models to test whether the 251-cluster structure and the
60.4\% optimization gap are an artifact of \texttt{all-mpnet-base-v2} or
generalize across embedding choices;
and (3)~\emph{description-generation LLM}---regenerating functional
descriptions with additional open-source models beyond Qwen2.5-30B-Instruct
and measuring the resulting cluster agreement (e.g., via Adjusted Rand Index)
to quantify how sensitive the semantic grouping is to the specific LLM used.
This ablation will let us report, rather than assume, the stability of our
clustering methodology across algorithmic and model choices.

\section{Conclusion}
\label{sec:conclusion}

We presented an automated pipeline that quantifies Dockerfile security technical
debt in enterprise inner-source repositories and identifies high-quality
configurations as remediation targets. Applied to 11,470 Dockerfiles from a
single large industrial company, the pipeline reveals a \textbf{60.4\% relative
improvement} in security posture achievable without developing new
templates---simply by propagating internal best practices that already exist
within each functional cluster. A conservative P90 target still yields a 28.75\%
global improvement.
As with any single-organization study, the exact magnitude of these figures
reflects this company's specific technology mix, governance posture, and
technical-debt distribution; other enterprises should expect the qualitative
pattern---latent internal references outnumbering the need for newly authored
templates---to recur, but the precise percentages are not a priori transferable
without a comparable study conducted in their own inner-source ecosystem.

The three research questions are answered with actionable precision.
\textbf{RQ1}: 99\% misconfiguration rate, 80.8\% smell prevalence, median
recency of 838~days, and a mean score of~0.48 establish a systemic, pervasive
deficit that is empirically documented for the first time in a corporate
inner-source context.
\textbf{RQ2}: LLM-semantic clustering reduces global base-image entropy
by~62\% with only 17\% noise, outperforming content-based and HLS syntactic
baselines in functional cohesion; the force-directed graph confirms workload
isolation across six distinct technology stacks.
\textbf{RQ3}: 60.4\% of measured debt is addressable from existing internal
data without commissioning new secure templates.

These findings provide infrastructure teams with a data-driven prioritization
framework, establish the enriched dataset as a retrieval corpus for
LLM-based automated repair, and quantify---for the first time in an enterprise
context---the scale of container security debt that automated software
engineering can resolve.

\section*{Data Availability Statement}
\label{sec:data}

The dataset analyzed in this study consists of Dockerfiles and associated metadata collected
from a proprietary enterprise inner-source environment.
Due to licensing restrictions and confidentiality agreements, these artifacts cannot be released
publicly or shared with external entities.
The inner-source license governing these repositories explicitly prohibits external distribution.
As a result, the dataset that led to the results in this paper is not available for review or public access.

The analysis pipeline is likewise not publicly released in source-code form.
The implementation contains proprietary integrations with an internal GitLab instance,
company-specific credential management, and internal toolchain orchestration that are
subject to corporate intellectual property restrictions.
To support replication in other enterprise contexts, the complete methodology---including
all hyperparameters, scoring formulas, prompt template, normalization thresholds,
and algorithmic choices---is fully specified in Section~\ref{sec:approach},
enabling independent reimplementation without access to the original codebase.

\begin{acks}
The authors used GitHub Copilot and Claude (Anthropic) to assist with language
improvement and document structuring during the preparation of this manuscript.
All scientific content, results, and conclusions are solely the work of the
authors.
\end{acks}

\bibliographystyle{ACM-Reference-Format}
\bibliography{main}

@inproceedings{cito_empirical_2017,
  author    = {Cito, J\"{u}rgen and Schermann, Gerald and Wittern, John Erik and
               Leitner, Philipp and Zumberi, Sali and Gall, Harald C.},
  title     = {An Empirical Analysis of the {Docker} Container Ecosystem on {GitHub}},
  booktitle = {Proceedings of the 14th International Conference on Mining Software
               Repositories (MSR)},
  year      = {2017},
  pages     = {323--333},
  doi       = {10.1109/MSR.2017.67},
  publisher = {IEEE},
  address   = {Buenos Aires, Argentina}
}

@inproceedings{eng_revisiting_2021,
  author    = {Eng, Kalvin and Hindle, Abram},
  title     = {Revisiting {Dockerfiles} in Open Source Software Over Time},
  booktitle = {Proceedings of the 18th International Conference on Mining Software
               Repositories (MSR)},
  year      = {2021},
  pages     = {449--459},
  doi       = {10.1109/MSR52588.2021.00057},
  publisher = {IEEE},
  address   = {Madrid, Spain}
}

@article{wu_characterizing_2020,
  author  = {Wu, Yiwen and Zhang, Yang and Wang, Tao and Wang, Huaimin},
  title   = {Characterizing the Occurrence of {Dockerfile} Smells in Open-Source
             Software: An Empirical Study},
  journal = {IEEE Access},
  year    = {2020},
  volume  = {8},
  pages   = {34127--34139},
  doi     = {10.1109/ACCESS.2020.2973750}
}

@inproceedings{lin_large-scale_2020,
  author    = {Lin, Changyuan and Nadi, Sarah and Khazaei, Hamzeh},
  title     = {A Large-scale Data Set and an Empirical Study of {Docker} Images
               Hosted on {Docker Hub}},
  booktitle = {Proceedings of the IEEE International Conference on Software
               Maintenance and Evolution (ICSME)},
  year      = {2020},
  pages     = {371--381},
  doi       = {10.1109/ICSME46990.2020.00043},
  publisher = {IEEE},
  address   = {Adelaide, Australia}
}

@inproceedings{haque_well_2022,
  author    = {Haque, Mubin Ul and Babar, M.~Ali},
  title     = {Well Begun is Half Done: An Empirical Study of Exploitability \&
               Impact of Base-Image Vulnerabilities},
  booktitle = {Proceedings of the IEEE International Conference on Software
               Analysis, Evolution and Reengineering (SANER)},
  year      = {2022},
  pages     = {1066--1077},
  doi       = {10.1109/SANER53432.2022.00124},
  publisher = {IEEE},
  address   = {Honolulu, HI, USA}
}

@inproceedings{opdebeeck_docker_2023,
  author    = {Opdebeeck, Ruben and Lesy, Jonas and Zerouali, Ahmed and
               De~Roover, Coen},
  title     = {The {Docker Hub} Image Inheritance Network: Construction and
               Empirical Insights},
  booktitle = {Proceedings of the 23rd International Working Conference on
               Source Code Analysis and Manipulation (SCAM)},
  year      = {2023},
  pages     = {198--208},
  doi       = {10.1109/SCAM59687.2023.00029},
  publisher = {IEEE},
  address   = {Bogot\'{a}, Colombia}
}

@inproceedings{shi_dr_2025,
  author    = {Shi, Hequan and Ying, Lingyun and Chen, Libo and Duan, Haixin
               and Liu, Ming and Xue, Zhi},
  title     = {Dr. {Docker}: A Large-Scale Security Measurement of {Docker}
               Image Ecosystem},
  booktitle = {Proceedings of the ACM on Web Conference (WWW)},
  year      = {2025},
  pages     = {2813--2823},
  doi       = {10.1145/3696410.3714653},
  publisher = {ACM},
  address   = {Sydney, Australia}
}

@article{rosa_mining_2025,
  author  = {Rosa, Giovanni and Guglielmi, Emanuela and Iannone, Mattia and
             Scalabrino, Simone and Oliveto, Rocco},
  title   = {Mining and measuring the impact of change patterns for improving
             the size and build time of {Docker} images},
  journal = {Empirical Software Engineering},
  year    = {2025},
  volume  = {30},
  number  = {5},
  pages   = {150},
  doi     = {10.1007/s10664-025-10680-8}
}

@inproceedings{bui_dockercleaner_2023,
  author    = {Bui, Quang-Cuong and Lauk\"otter, Malte and Scandariato, Riccardo},
  title     = {{DockerCleaner}: Automatic Repair of Security Smells in
               {Dockerfiles}},
  booktitle = {Proceedings of the IEEE International Conference on Software
               Maintenance and Evolution (ICSME)},
  year      = {2023},
  pages     = {160--170},
  doi       = {10.1109/ICSME58846.2023.00026},
  publisher = {IEEE},
  address   = {Bogot\'{a}, Colombia}
}

@inproceedings{durieux_empirical_2024,
  author    = {Durieux, Thomas},
  title     = {Empirical Study of the {Docker} Smells Impact on the Image Size},
  booktitle = {Proceedings of the IEEE/ACM 46th International Conference on
               Software Engineering (ICSE)},
  year      = {2024},
  pages     = {1--12},
  doi       = {10.1145/3597503.3639143},
  publisher = {ACM},
  address   = {Lisbon, Portugal}
}

@inproceedings{ksontini_refactoring_2025,
  author    = {Ksontini, Emna and Mastouri, Meriem and Khalsi, Rania and
               Kessentini, Wael},
  title     = {Refactoring for {Dockerfile} Quality: A Dive into Developer
               Practices and Automation Potential},
  booktitle = {Proceedings of the 22nd International Conference on Mining
               Software Repositories (MSR)},
  year      = {2025},
  pages     = {788--800},
  doi       = {10.1109/MSR66628.2025.00116},
  publisher = {IEEE},
  address   = {Ottawa, ON, Canada}
}

@inproceedings{ksontini_refactorings_2021,
  author    = {Ksontini, Emna and Kessentini, Marouane and Ferreira, Thiago~do~N.
               and Hassan, Foyzul},
  title     = {Refactorings and Technical Debt in {Docker} Projects: An
               Empirical Study},
  booktitle = {Proceedings of the 36th IEEE/ACM International Conference on
               Automated Software Engineering (ASE)},
  year      = {2021},
  pages     = {781--791},
  doi       = {10.1109/ASE51524.2021.9678585},
  publisher = {IEEE},
  address   = {Melbourne, Australia}
}

@inproceedings{ye_llmsecconfig_2025,
  author    = {Ye, Ziyang and Le, Triet Huynh Minh and Babar, M.~Ali},
  title     = {{LLMSecConfig}: An {LLM}-Based Approach for Fixing Software
               Container Misconfigurations},
  booktitle = {Proceedings of the 22nd International Conference on Mining
               Software Repositories (MSR)},
  year      = {2025},
  pages     = {629--641},
  doi       = {10.1109/MSR66628.2025.00099},
  publisher = {IEEE},
  address   = {Ottawa, ON, Canada}
}

@inproceedings{rosa_automatically_2023,
  author    = {Rosa, Giovanni and Mastropaolo, Antonio and Scalabrino, Simone
               and Bavota, Gabriele and Oliveto, Rocco},
  title     = {Automatically Generating {Dockerfiles} via Deep Learning:
               Challenges and Promises},
  booktitle = {Proceedings of the IEEE/ACM International Conference on
               Software and System Processes (ICSSP)},
  year      = {2023},
  pages     = {1--12},
  doi       = {10.1109/ICSSP59042.2023.00011},
  publisher = {IEEE},
  address   = {Melbourne, Australia}
}

@article{rosa_fixing_2024,
  author  = {Rosa, Giovanni and Zappone, Federico and Scalabrino, Simone and
             Oliveto, Rocco},
  title   = {Fixing {Dockerfile} smells: an empirical study},
  journal = {Empirical Software Engineering},
  year    = {2024},
  volume  = {29},
  number  = {5},
  pages   = {108},
  doi     = {10.1007/s10664-024-10471-7}
}

@inproceedings{zhang_recommending_2022,
  author    = {Zhang, Yinyuan and Zhang, Yang and Mao, Xinjun and Wu, Yiwen and
               Lin, Bo and Wang, Shangwen},
  title     = {Recommending Base Image for {Docker} Containers based on Deep
               Configuration Comprehension},
  booktitle = {Proceedings of the IEEE International Conference on Software
               Analysis, Evolution and Reengineering (SANER)},
  year      = {2022},
  pages     = {449--453},
  doi       = {10.1109/SANER53432.2022.00060},
  publisher = {IEEE},
  address   = {Honolulu, HI, USA}
}

@inproceedings{campello2013density,
  author    = {Campello, Ricardo J.~G.~B. and Moulavi, Davoud and Sander,
               J\"{o}rg},
  title     = {Density-based Clustering Based on Hierarchical Density
               Estimates},
  booktitle = {Advances in Knowledge Discovery and Data Mining (PAKDD)},
  year      = {2013},
  pages     = {160--172},
  publisher = {Springer},
  address   = {Gold Coast, Australia},
  doi       = {10.1007/978-3-642-37456-2_14},
  url       = {https://doi.org/10.1007/978-3-642-37456-2_14}
}

@techreport{cncf_survey_2024,
  author      = {{Cloud Native Computing Foundation}},
  title       = {{CNCF} Annual Survey 2024},
  year        = {2024},
  institution = {Cloud Native Computing Foundation},
  type        = {Industry Report},
  url         = {https://www.cncf.io/reports/cncf-annual-survey-2024/}
}

@inproceedings{10.5555/3290281.3290310,
author = {Bank, Erin and Gr\"{u}tter, Georg and Hanmer, Robert and Stol, Klaas-Jan and Sudarsan, Padma and Williams, Cedric and Yao, Tim and Yeates, Nick},
title = {Innersource patterns for collaboration},
year = {2017},
isbn = {9781941652060},
publisher = {The Hillside Group},
address = {USA},
booktitle = {Proceedings of the 24th Conference on Pattern Languages of Programs},
articleno = {24},
numpages = {15},
location = {Vancouver, British Columbia, Canada},
series = {PLoP '17},
doi = {10.5555/3290281.3290310},
url = {https://dl.acm.org/doi/abs/10.5555/3290281.3290310}
}

@article{10.1145/3611648,
author = {Buchner, Stefan and Riehle, Dirk},
title = {The Business Impact of Inner Source and How to Quantify It},
year = {2023},
issue_date = {February 2024},
publisher = {Association for Computing Machinery},
address = {New York, NY, USA},
volume = {56},
number = {2},
issn = {0360-0300},
url = {https://doi.org/10.1145/3611648},
doi = {10.1145/3611648},
journal = {ACM Comput. Surv.},
month = sep,
articleno = {47},
numpages = {27}
}

@inproceedings{shabani2025flakiness,
author = {Shabani, Taha and Nashid, Noor and Alian, Parsa and Mesbah, Ali},
title = {Dockerfile Flakiness: Characterization and Repair},
year = {2025},
isbn = {9798331505691},
publisher = {IEEE Press},
doi = {10.1109/ICSE55347.2025.00238},
url = {https://doi.org/10.1109/ICSE55347.2025.00238},
booktitle = {Proceedings of the IEEE/ACM 47th International Conference on Software Engineering},
pages     = {1793--1805},
numpages  = {13},
address   = {Ottawa, ON, Canada}
}

\end{document}